# Patterning of high mobility electron gases at complex oxide interfaces

F. Trier,[1,a] G. E. D. K. Prawiroatmodjo,[2] M. von Soosten,[2] D. V. Christensen,[1] T. S. Jespersen,[2] Y. Z. Chen,[1] and N. Pryds,[1]

[1]*Department of Energy Conversion and Storage, Technical University of Denmark, Risø Campus, DK-4000 Roskilde, Denmark*

[2]*Niels Bohr Institute, Center for Quantum Devices, University of Copenhagen, DK-2100 Copenhagen Ø, Denmark*

a) Corresponding author – fetri@dtu.dk

Oxide interfaces provide an opportunity for electronics. However, patterning of electron gases at complex oxide interfaces is challenging. In particular, patterning of complex oxides while preserving a high electron mobility remains underexplored and inhibits the study of quantum mechanical effects where extended electron mean free paths are paramount. This letter presents an effective patterning strategy of both the amorphous-LaAlO$_3$/SrTiO$_3$ (a-LAO/STO) and modulation-doped amorphous-LaAlO$_3$/La$_{7/8}$Sr$_{1/8}$MnO$_3$/SrTiO$_3$ (a-LAO/LSM/STO) oxide interfaces. Our patterning is based on selective wet etching of amorphous-LSM (a-LSM) thin films which acts as a hard mask during subsequent depositions. Strikingly, the patterned modulation-doped interface shows electron mobilities up to ~8,700 cm$^2$/Vs at 2 K, which is among the highest reported values for patterned conducting complex oxide interfaces that usually are ~1,000 cm$^2$/Vs at 2 K.

Research on interface phenomena between the two insulating oxides LaAlO$_3$ (LAO) and SrTiO$_3$ (STO) have resulted in the discovery of a wealth of attractive properties. These include a metallic two-dimensional electron gas (2-DEG),[1] an electric field controlled metal-insulator transition[2,3], a superconducting phase[4] tunable by application of an electric field,[5,6] and ferromagnetic ordering[7]. Herein, the charge carrier mobility represents the Achilles tendon of the interface conductivity as this generally tends to be low for complex oxides. Therefore, with the aim of improving the performance of oxide electronics[8,9] and to allow the study of quantum mechanical effects where extended electron mean free paths are paramount, it is necessary to enhance the inherent charge carrier mobility. This has seen significant progress with heterostructures of spinel structured γ-Al$_2$O$_3$ on STO (GAO/STO) displaying record-high electron mobilities of up to 140.000 cm$^2$/Vs at 2 K.[10] Recently, another high electron mobility system was discovered by introducing a single unit cell La$_{7/8}$Sr$_{1/8}$MnO$_3$ (LSM) spacer layer between STO and amorphous-LAO grown at room temperature (RT).[11] Strikingly, this system showed enhanced electron mobilities of up to 70,000 cm$^2$/Vs at 2 K compared to the usual ~1,000 cm$^2$/Vs at 2 K for the a-LAO/STO heterostructure.[11,12]

An equally important element inhibiting the advancement of oxide electronics and the ability to study quantum mechanical phenomena at mesoscopic scales is a lithographic patterning scheme which preserves the inherent interface quality of the system in spite of the processing. i.e. a strategy for patterning of the existing record-high mobility electron gases such as those found in GAO/STO or a-LAO/LSM/STO. Patterning of complex oxides has previously been addressed either relying on hard mask lift-off[13,14,15] or low-energy ion beam irradiation.[16] However, we found that these methods were challenging in patterning the high mobility GAO/STO heterostructure prepared at high temperature.[10]

Although high mobility oxide interfaces prepared at RT[11,17] provides a straightforward way to pattern oxide interfaces with conventional lithography techniques utilizing e.g. a resist soft mask, however, this results in insulating interfaces. Ultimately, state-of-the-art values of the electron mobility in patterned complex oxide interfaces typically remains around 3000 cm$^2$/Vs at 2 K.[14,17]

In this letter, we present a strategy which allows patterning of not only the a-LAO/STO interface conductivity but also the high mobility interface conductivity in the modulation-doped a-LAO/LSM/STO structure. To achieve this we initially covered the bare STO surface with an amorphous-LSM (a-LSM) thin film grown at RT (see Fig. 1). Here, a-LSM is chosen primarily since a-LSM/STO heterostructures inherently are insulating[12] regardless of temperature or oxygen partial pressure during the sample processing. During deposition of the LSM spacer layer ($P_{O2}$≈1×10$^{-4}$, T≈600 °C) the use of other potential hard mask materials such as amorphous aluminum oxide (AlO$_x$)[14] will be problematic as this could result in conducting AlO$_x$/STO interfaces[10,12,17] and thus prevent patterning of the interface conductivity. Furthermore, a-LSM is chosen since by selective wet chemical etching[18] it can act as a hard shadow mask during the subsequent depositions. With a final deposition of a-LAO, this will result in conducting a-LAO/STO areas whereas a-LAO/a-LSM/STO regions remain insulating (see Fig. 1(f)) owing to their different redox-reactivity with the STO substrate.[12] Remarkably, complex oxide Hall bar devices prepared with this patterning strategy show electron mobilities up to ~8,700 cm$^2$/Vs at 2 K.

Deposition of oxide thin films at RT permits the usage of soft resist masks which otherwise would decompose at elevated deposition temperatures. However, directly patterning the a-LAO/STO interface by use of a polymethyl methacrylate (PMMA) resist layer as a soft mask results in insulating interfaces (data not shown). To prevent the resist layer from contaminating the delicate STO surface, we investigated a hard mask patterning strategy: the TiO$_2$-terminated STO[19] substrates are initially deposited with 60 nm of a-LSM using pulsed laser deposition (PLD) (see Fig. 1(b)). The a-LSM/STO samples are then prepared with a 200 nm thick PMMA electron-beam resist layer which is exposed using a 100 kV electron-beam into the desired Hall bar geometry (see Fig. 1(c)) and developed using 1:3 methyl isobutyl ketone and isopropanol (MIBK:IPA). To improve the resist adhesion with the a-LSM surface the PMMA resist is reflown by post-baking at 185 °C for 90 s. The remaining resist will then protect underlying a-LSM from the etchant – a 2:2:35 KI(3M):HCl(35%):H$_2$O acid solution.[18] The a-LSM/STO samples are etched for 15 s at a temperature of 20 °C (see Fig. 1 (d)). This procedure allows pattern transfer with sub-micrometer resolution. To investigate the surface quality after etching, the sample surface at etched regions is probed using atomic force microscopy (AFM) (see Fig. 2(b)). For all measured samples, the STO surface structure is consistent with atomically flat TiO$_2$ terraces that have a width similar to what is measured prior to the etching process. After the remaining resist is removed (see Fig. 1(e)) the structured a-LSM/STO samples are transferred back to the PLD chamber where either a-LAO or a-LAO/LSM are deposited (see Fig. 1(f)). For the a-LAO/STO samples, 16 nm a-LAO is deposited on the structured a-LSM/STO sample at RT with identical PLD parameters as previously reported.[3,12] For the a-LAO/LSM/STO samples, the structured a-LSM/STO sample is initially deposited with a single unit cell LSM spacer layer at 600 ºC and otherwise identical deposition parameters as recently reported.[11] Subsequently, it is cooled under an oxygen pressure of $P_{O2}\approx 1\times 10^{-4}$ mbar with a rate of 15 ºC/min to RT (<25 ºC) followed by deposition of 16 nm d-LAO using the PLD parameters reported elsewhere.[3,12] After final a-LAO deposition, the Hall bar devices are imaged using optical microscopy (see Fig. 2(d)), where light and dark grey regions correspond to areas with and without the a-LSM hard mask, respectively. This visible difference between the two areas allows for easy localization of the Hall bar devices. Additionally, Fig. 2(e) shows the device topography at two opposing voltage probes as imaged by AFM. The device topography is well defined and displays sub-micrometer pattern-edge roughness.

For comparison, unpatterned 5×5 mm$^2$ a-LAO/STO and a-LAO/LSM/STO reference samples (i.e. without a-LSM deposition or etching) are prepared and measured in the Van der Pauw geometry. The interface of all samples is contacted using ultrasonically wire-bonded aluminum wires.

Fig. 3(a) shows the sheet resistance as a function of temperature for a representative a-LAO/STO Hall bar and the unpatterned a-LAO/STO reference sample. Both interfaces show comparable transport properties, indicating that the interface conduction generally is little affected by the patterning process. The a-LAO/STO Hall bar albeit displays a slightly higher sheet resistance than the unpatterned a-LAO/STO at room temperature but this difference diminishes as the temperature is decreased. This discrepancy is caused by their minor carrier density difference for $T > 100$ K (see Fig. 3(b)) below which the cubic to tetragonal phase transition of STO occurs.[20] Finally, as shown in Fig. 3(c), the two a-LAO/STO samples exhibit almost same electron mobilities, indicating that the quality/cleanness of the interface is preserved after the patterning process. Similarly, the mobility is comparable with typical values for patterned and unpatterned interface conductivity in most all-crystalline LAO/STO heterostructures.[13,15]

Interestingly, this strategy is also applicable to pattern modulation-doped a-LAO/LSM/STO Hall bar devices with enlarged electron mobilities. As shown in Fig. 4, a patterned a-LAO/LSM/STO Hall bar device shows a carrier density of $5.6\times 10^{12}$ cm$^{-2}$ (see Fig. 4(b)), much lower than the a-LAO/STO samples (see Fig. 3(b)). Moreover, the carrier density is little temperature dependent similarly to what is characteristic for the unpatterned a-LAO/LSM/STO heterostructure. Strikingly, the patterned a-LAO/LSM/STO Hall bar device shows a mobility of 8,703 cm$^2$/Vs at 2 K (see Fig. 4(c)). Although this mobility remains almost an order below the record-high value of 70,000 cm$^2$/Vs for the unpatterned a-LAO/LSM/STO reference sample, it is among the highest reported values for patterned complex oxide interfaces with the typical value for the patterned LAO/STO interface often being below ~1000 cm$^2$/Vs at 2 K.[13,15] Such a high mobility and low carrier density interface in a-LAO/LSM/STO enables the observation of clear Shubnikov-de Haas oscillations and the initial manifestation of the quantum Hall effect in complex oxides.[11]

To conclude, we outline a general strategy for patterning of metallic interfaces in complex oxide heterostructures prepared at RT. In particular, the strategy is based on selective etching of an a-LSM thin film acting as hard mask in subsequent film depositions. The technique is further found to be applicable for the modulation-doped oxide interface where the patterned interfaces show enhanced electron mobilities compared to typical values of the canonical LAO/STO interface. This opens the door to design oxide microelectronic devices and study mesoscopic physics based on complex oxides.

*Figure captions*

FIG. 1. (Color online) Schematic illustration of the patterning process. (a,b) A bare $TiO_2$-terminated $SrTiO_3$ (STO) substrate is deposited with amorphous-$La_{7/8}Sr_{1/8}MnO_3$ (a-LSM/STO). (c) The a-LSM/STO heterostructure is then prepared with an electron-beam (e-beam) defined resist pattern. (d) The sample is then subjected to selective KI/HCl etching as directed by the resist. (e) The remaining resist is then removed and the sample surface is cleaned. (f) Deposition of amorphous-$LaAlO_3$ (a-LAO) results in either conducting a-LAO/STO or insulating a-LAO/a-LSM/STO heterostructures. The high electron mobility interface is obtained by deposition of a single unit cell LSM before the a-LAO deposition.

FIG. 2. (Color online) (a) Schematic illustration of an etched a-LSM/STO heterostructure. The dashed square schematically represents the scanned area imaged by atomic force microscopy (AFM) in (b). (b) AFM image of a KI/HCl etched region with visible STO terrace structure. (c) Schematic illustration of the patterned a-LAO/STO heterostructure. (d) Optical microscopy image of a a-LAO/STO Hall bar device with light and dark grey regions corresponding to areas with d-LAO/STO and a-LAO/a-LSM/STO, respectively. The dashed square indicates the AFM scanned area in (e). (e) AFM image of a Hall bar segment and two voltage probes which shows sub-micrometer pattern-edge roughness.

FIG. 3. (Color online) (a) Sheet resistance ($R_{xx}$) vs. temperature ($T$) of a representative patterned a-LAO/STO Hall bar with a width of 50 μm and a distance between longitudinal voltage probes of 300 μm and the unpatterned a-LAO/STO reference sample which indicates that the patterning process has little effect on the formation of interface conductivity. (b) Carrier density ($n_s$) as a function of $T$ shows that there is a small difference between the two samples for $T > 100$ K. (c) Comparing the electron mobility (μ) of the two samples shows that the quality of the interface is preserved in spite of the patterning process.

FIG. 4. (Color online) (a) $T$ dependence of $R_{xx}$ for a representative patterned a-LAO/LSM/STO Hall bar with a width of 70 μm and a distance between longitudinal voltage probes of 300 μm and the unpatterned a-LAO/LSM/STO reference sample. (b) $n_s$ as a function of $T$ indicates a carrier density difference between the patterned and unpatterned samples which is consistent with their $R_{xx}$ discrepancy. (c) At low $T$, the electron mobility, μ, of the two samples are consistent with high quality interfaces, and for the patterned sample the measured μ is among the highest reported values for patterned complex oxide interfaces.

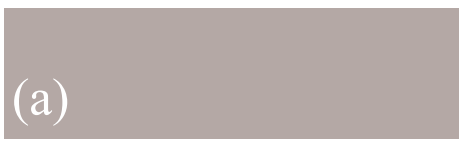
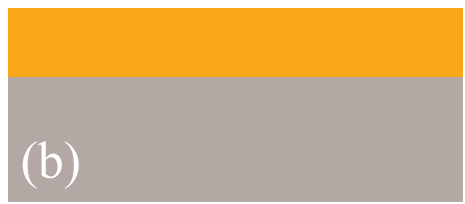
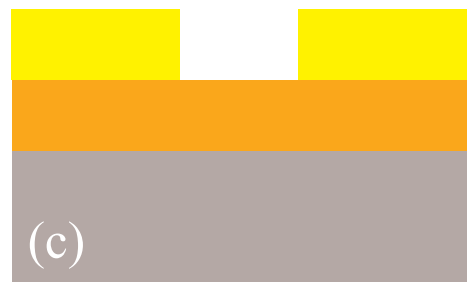
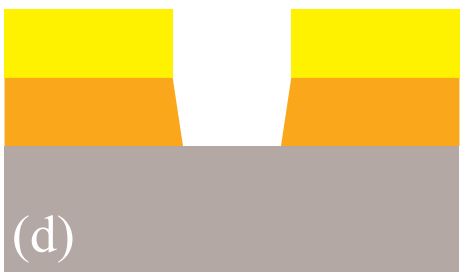
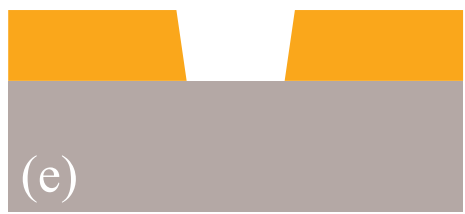
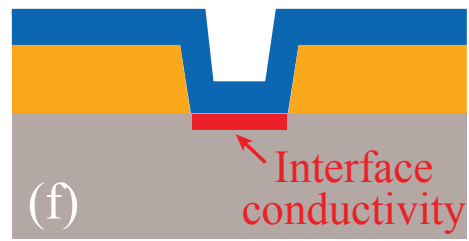

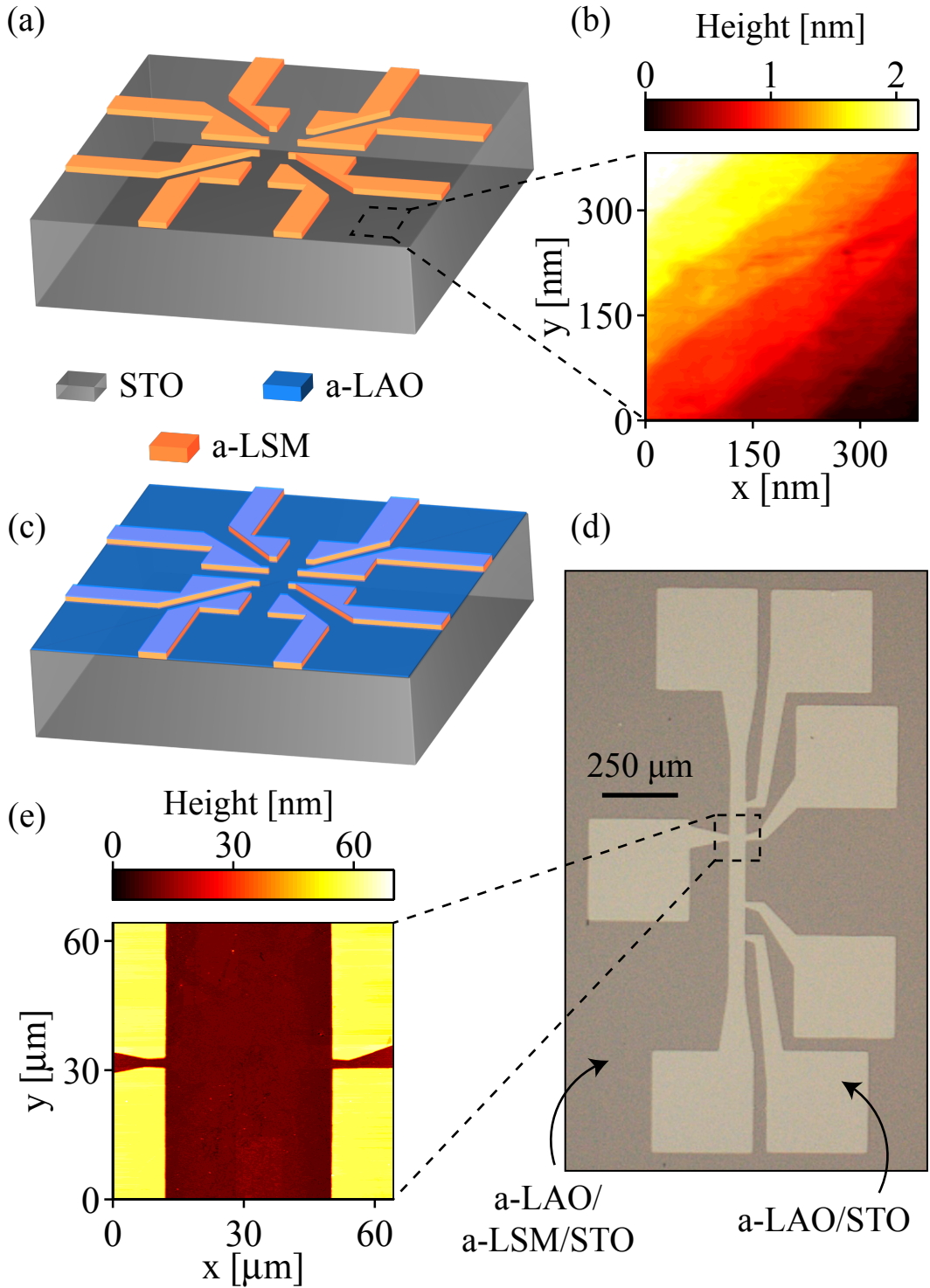

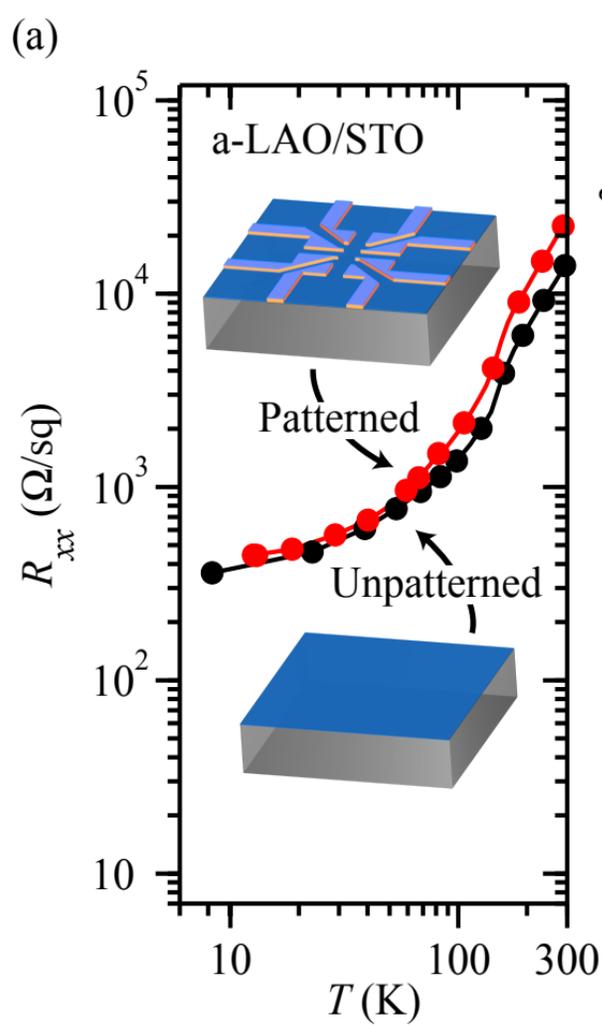

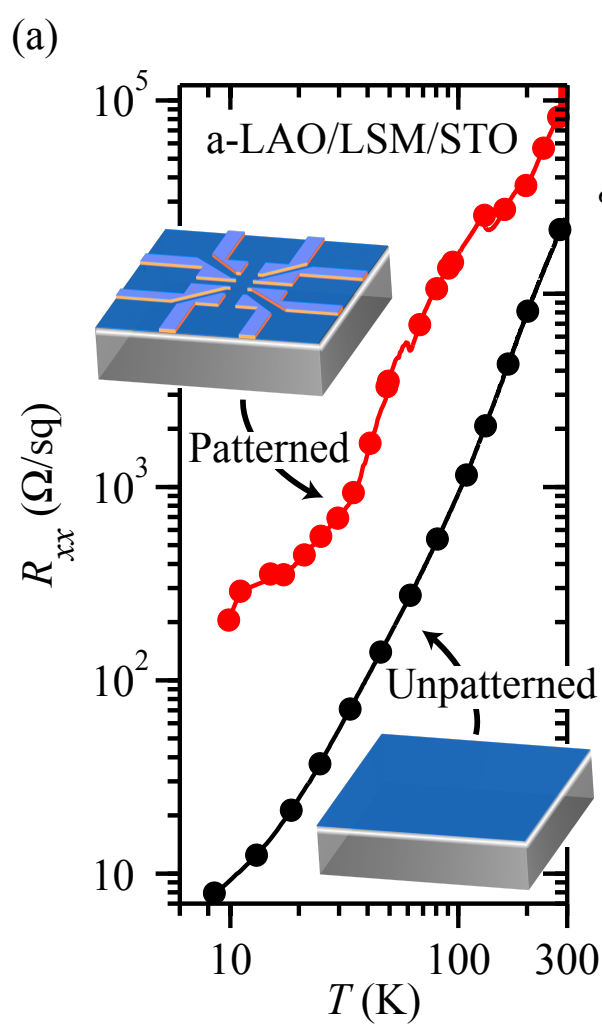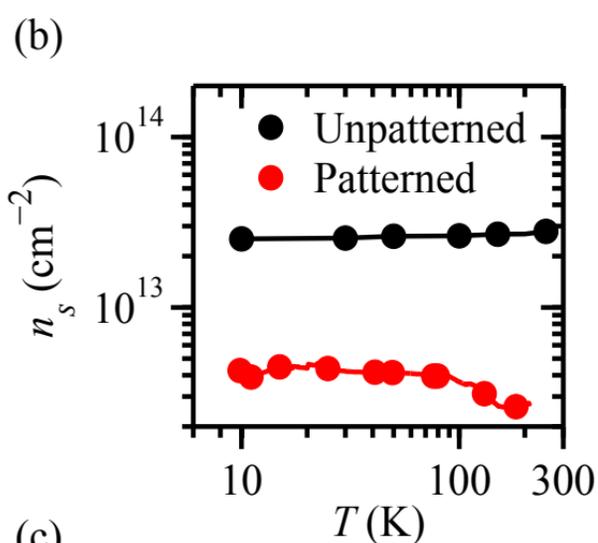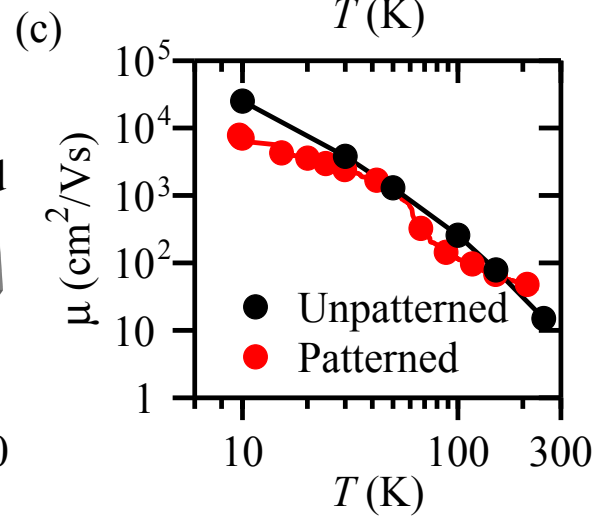